\documentclass[11pt,a4paper]{article}

\usepackage{amsmath,mathtools,amssymb,amsfonts,bbm}
\usepackage[top=3.11cm,bottom=4.11cm,right=3.11cm,left=3.11cm]{geometry}
\usepackage{physics}
\usepackage{slashed}
\usepackage{xcolor,graphicx}
\usepackage[export]{adjustbox}
\usepackage{subfigure}
\usepackage{etoolbox}
\usepackage{bbold}
\usepackage{xspace}
\usepackage{appendix}
\usepackage{amsthm}
\usepackage{makecell}

\usepackage[utf8]{inputenc}
\usepackage[T1]{fontenc}
\usepackage[UKenglish]{babel}

\usepackage{tabularx}
\newcolumntype{C}{>{\centering\arraybackslash}X}
\newcolumntype{R}{>{\raggedleft\arraybackslash}X}
\newcolumntype{L}{>{\raggedright\arraybackslash}X}

\usepackage{multirow, bigdelim}

\usepackage[style=ieee, citestyle=numeric-comp, backend=biber]{biblatex}
\numberwithin{equation}{section}
\usepackage[hidelinks]{hyperref} 
\usepackage[capitalise]{cleveref}

\usepackage{xurl}
\hypersetup{breaklinks=true}

\allowdisplaybreaks

\newcommand{\intx}{\int d^4x}
\newcommand{\Dintx}{\int d^Dx}

\newcommand{\projR}{\mathbb{P}_{\mathrm{R}}}
\newcommand{\projL}{\mathbb{P}_{\mathrm{L}}}

\theoremstyle{definition}

\title{Two-Loop Renormalization of a Chiral $SU(2)$ Gauge Theory in Dimensional Regularization with Non-Anticommuting $\gamma_5$}

\newcommand{\email}{}
\author{
Paul Kühler,$^a$\thanks{\email{paul.kuehler@tu-dresden.de}}\quad
Dominik  Stöckinger,$^a$\thanks{\email{dominik.stoeckinger@tu-dresden.de}}\\[2em]
$^a$Institut für Kern- und Teilchenphysik, TU Dresden,\\ Zellescher Weg 19, DE-01069 Dresden, Germany}

\begin{document}
\thispagestyle{empty}

\maketitle

\setcounter{footnote}{0}
\vspace{2ex}
\begin{abstract}
\noindent
Higher order calculations in chiral gauge theories such as the Electroweak Standard Model require a sound treatment of the notoriously problematic $\gamma_5$-matrix in Dimensional Regularization (DReg). In the all-order consistent BMHV scheme anticommutativity has to be sacrificed, resulting in spurious breakings of BRST invariance, the restoration of which necessitates finite, symmetry-restoring counterterms. Following recent advances in successfully applying this scheme to multi-loop calculations for Abelian models, we shall here present the first complete non-Abelian two-loop result for the case of $SU(2)$, which is of particular interest to the Standard Model. We provide the complete list of finite, two-loop symmetry restoring counterterms and discuss intricacies of the non-Abelian implementation.
Except for one novel term, the finite counterterm action exhibits the same structure as at one-loop order.
\end{abstract}

\newpage
\setcounter{page}{1}

\tableofcontents\newpage


\section{Introduction}
High-precision tests of the Standard Model (SM) are a promising and important area in elementary particle physics. They call for high-precision theoretical predictions involving multi-loop perturbative calculations. Dimensional Regularization (DReg) is a particularly convenient regularization since it manifestly preserves Lorentz invariance and allows for efficient computational techniques. It also preserves gauge and BRST invariance, albeit only in vector-like gauge theories such as QED or QCD.

For the high-precision calculations required today and in the future, taking into account electroweak corrections is of increasing importance \cite{Blondel:2018mad}. Unlike QED and QCD, the electroweak sector of the SM (EWSM) is a chiral gauge theory, where left- and right-handed fermions couple differently to the gauge bosons. In this context, DReg leads to the well known $\gamma_5$ problem \cite{tHooft:1972tcz}, which implies that couplings of chiral fermions to gauge fields cannot be consistently continued to formally $D$ dimensions without giving up important properties. Many proposals for treatments exists \cite{Jegerlehner:2000dz, Kreimer:1989ke,Chanowitz:1979zu,Korner:1991sx,Zerf:2019ynn,Chen:2023lus,Chen:2024zju}, but the scheme defined by 't Hooft, Veltman and Breitenlohner, Maison (BMHV) \cite{tHooft:1972tcz,Breitenlohner:1975hg,Breitenlohner:1976te,Breitenlohner:1977hr} stands out as rigorous, with established all-order theorems on its renormalization properties and internal consistency.
In the BMHV scheme, the anticommutativity of $\gamma_5$ is given up; $\gamma_5$ only anticommutes with $\bar{\gamma}^\mu$ matrices corresponding to the original 4-dimensional space, while it commutes with the evanescent $\hat{\gamma}^\mu$ matrices corresponding to the remaining $(D-4)$ dimensions. It turns out that as a result BRST symmetry of chiral gauge theories is necessarily broken in $D$ dimensions, and it is required to add appropriate symmetry-restoring counterterms which restore the symmetry after renormalization in the 4-dimensional limit, for early results see Refs.~\cite{Jones:1982zf, Freitas:2002ja, Martin:1999cc,Sanchez-Ruiz:2002pcf}, for a recent review see e.g.~\cite{Belusca-Maito:2023wah} and for a compact summary of recent progress see \cite{Belusca-Maito:2022usw,Kuhler:2024fak}.

In recent years, the BMHV scheme has gained in popularity and progress has been achieved in the context of renormalizable gauge theories \cite{Martin:1999cc,Sanchez-Ruiz:2002pcf,Belusca-Maito:2020ala,Belusca-Maito:2021lnk,Belusca-Maito:2022wem,Stockinger:2023ndm,Cornella:2022hkc,OlgosoRuiz:2024dzq,Ebert:2024xpy} and effective field theories \cite{Carmona:2021xtq,Fuentes-Martin:2022vvu,DiNoi:2023ygk,Naterop:2023dek,Naterop:2024ydo,DiNoi:2025uan}. Specifically in the case of renormalizable gauge theories Refs.~\cite{Martin:1999cc,Sanchez-Ruiz:2002pcf,Belusca-Maito:2020ala,Belusca-Maito:2021lnk,Belusca-Maito:2022wem,Stockinger:2023ndm,Cornella:2022hkc,OlgosoRuiz:2024dzq,Ebert:2024xpy}
have evaluated the spurious symmetry breaking and determined required symmetry-restoring counterterms in various models. The complexity of the studies depends on three kinds of properties: the loop order, the gauge group, and the types of allowed interactions. Earlier studies \cite{Martin:1999cc,Sanchez-Ruiz:2002pcf,Belusca-Maito:2020ala,Belusca-Maito:2021lnk,Belusca-Maito:2022wem,Stockinger:2023ndm} have allowed only interactions that do not mix left- and right-handed fermions on the evanescent level; Refs.~\cite{Cornella:2022hkc,OlgosoRuiz:2024dzq,Ebert:2024xpy} have successively allowed for more general interaction structures and studied the resulting impact on symmetry-restoring counterterms. Out of the mentioned works, only Refs.~\cite{Belusca-Maito:2021lnk,Stockinger:2023ndm} study renormalization and symmetry restoration at the two-loop and three-loop level, respectively. They show that, despite the additional complication caused by subdivergences and subrenormalization, symmetry restoration works by adding counterterms of a similar structure as at the 1-loop level.

However, so far all multi-loop studies of BRST symmetry restoration were restricted to Abelian gauge theories. In the present paper we present a first two-loop study for a non-Abelian gauge theory. To be concrete and motivated by the EWSM, we focus on an SU(2) gauge theory with a set of chiral, right-handed fermions that are allowed to form an arbitrary reducible representation. We carry out the complete two-loop renormalization procedure and determine the symmetry-restoring counterterms that restore the validity of the Slavnov-Taylor identity, and the UV divergent counterterms that render the theory finite. A key novel feature consists in the plethora of non-Abelian structures such as interacting ghosts and loop corrections to non-linear BRST transformations, many of which matter starting from the two-loop level. Another important difference to previous works are technical hurdles in the coherent {\tt FeynArts} implementation of required exotic quantities such as operators and external fields with non-vanishing ghost number and wrong statistics.

The paper is structured as follows. First, in Sec.~\ref{Sec:ModelDefinition} we define the model, its group structure and its continuation to $D$ dimensions in the BMHV scheme. Section \ref{Sec:DeterminationOfFCT} then focuses on the strategy to determine the symmetry breaking and symmetry restoration and describes the additional complications due to the non-Abelian structure and the multi-loop level in detail. Section \ref{Sec:CountertermResults}  lists the UV divergent two-loop counterterms and presents the main result of the paper, the complete list of finite symmetry-restoring counterterms at the two-loop level. Finally, Sec.~\ref{Sec:Conclusions} presents the conclusions.

\section{Chiral SU(2) model in $D$ dimensions}\label{Sec:ModelDefinition}
In this section we specify our model by defining the tree-level Lagrangian and BRST transformations in DReg as well as some group notation. Throughout the paper we use the customary notation for Lorentz covariants in the BMHV scheme of the earlier Refs.~\cite{Belusca-Maito:2020ala,Belusca-Maito:2021lnk, Stockinger:2023ndm,Ebert:2024xpy} and the review \cite{Belusca-Maito:2023wah}, where in the $D=4-2\epsilon$-dimensional space e.g.~$\gamma^{\mu}, \overline{\gamma}^{\mu}, \widehat{\gamma}^{\mu}$ denote $D, 4$- and $-2\epsilon$-dimensional covariants, respectively.  The $\epsilon$-dimensional hatted covariants are also called evanescent. For the $\gamma$-matrices and $\gamma_5$ we have
\begin{equation}
    \gamma^{\mu}=\overline{\gamma}^{\mu}+\widehat{\gamma}^{\mu},\qquad[\overline{\gamma}^{\mu},\gamma_5]=0,\qquad[\widehat{\gamma}^{\mu},\gamma_5]=2\widehat{\gamma}^{\mu}\gamma_5.
\end{equation}
\subsection{Group notation}
Our model has the generic form of a Yang-Mills theory analogously to the one studied in Ref.~\cite{Belusca-Maito:2020ala} at the one-loop level. For definiteness we fix the gauge group to SU(2), corresponding to the non-Abelian part of the electroweak SM gauge group. For the matter part we include $N_f$ right-handed fermions $\psi_{Ri}$ collected in one multiplet which transforms under a fully reducible representation $\mathcal{R}$ of $SU(2)$ with generators $T^a(\mathcal{R})_{ij}\equiv T^a_{\mathcal{R}ij}$ ($a\in\{1,\dots,3\}$), satisfying the SU(2) algebra $[T^a(\mathcal{R}),T^b(\mathcal{R})]=i\epsilon_{abc}T^c(\mathcal{R})$. 
The generators are block-diagonal, 
\begin{equation}
                T^a_{\mathcal{R}}=\mathrm{diag}(T^a_{\mathcal{R}_1}\dots T^a_{\mathcal{R}_{M}}),
\end{equation}
with the blocks corresponding to $M$ irreducible $SU(2)$-representations $\mathcal{R}_m$ ($m\in\{1,\dots,M\}$). The $N_f=\sum_{m}^M\mathrm{dim}(\mathcal{R}_m)$ fermions are grouped accordingly in $n$-plets ($n=\mathrm{dim}(\mathcal{R}_m)$) of the irreducible representations $\mathcal{R}_m$.
The normalization of the generators is given by
\begin{equation}
    \Tr_{\mathcal{R}}(T_{\mathcal{R}}^aT_{\mathcal{R}}^b)=\sum_{m}\Tr_{\mathcal{R}_m}(T^a_{\mathcal{R}_m}T^b_{\mathcal{R}_m})=\sum_{m}S_2(\mathcal{R}_m)\delta^{ab}\equiv\sum_{\mathcal{R}}S_2(\mathcal{R})\delta^{ab}.
\end{equation}
The $SU(2)$ group has only one Casimir operator per irreducible representation which is proportional to a unit matrix by Schur's lemma; hence for the fully reducible representation it becomes,
\begin{equation}
    (T^a_{\mathcal{R}}T^a_{\mathcal{R}})_{ij}=(C_2(\mathcal{R}))_{ij}=\mathrm{diag}(C_2(\mathcal{R}_1),\dots,C_2(\mathcal{R}_{M})),
\end{equation}
where $C_2(\mathcal{R}_m)_{ij}=C_2(\mathcal{R}_m)\delta^{(m)}_{ij}$ and ${i,j=1\dots m},\; C_2(\mathcal{R}_m)\in\mathbb{R}$.
The structure constant of $SU(2)$ is given by the $3$-dimensional fully antisymmetric epsilon tensor $\epsilon^{abc}$, which also defines the adjoint representation,
\begin{equation}
    T^a(\mathcal{A})_{bc}=-i(\epsilon^a)_{bc},
\end{equation}
with the corresponding Casimir operator,
\begin{equation}
    (T^a(\mathcal{A})T^a(\mathcal{A}))_{bc}=C_{\mathcal{A}}\delta^{bc}=2\delta^{bc}.
\end{equation}
We note in passing that the usual requirement on the matter content of any chiral gauge theory to be chosen such that the gauge anomaly vanishes here is fulfilled by construction since $SU(2)$ is free of perturbative anomalies owing to the vanishing of the fully-symmetric rank-$3$ tensor $d^{abc}_{\mathcal{R}_m}=0$ for any irreducible $\mathcal{R}_m$.

\subsection{Definition of the DReg Lagrangian and BRST breaking}
As discussed e.g.~in Refs.~\cite{Belusca-Maito:2020ala,Ebert:2024xpy} there is no unique extension of a 4-dimensional Lagrangian to $D$ dimensions. If the physical fermions are all right-handed, as here and in Ref.~\cite{Belusca-Maito:2020ala},  we have to additionally introduce sterile left-handed fields to render the kinetic fermion term fully $D$-dimensional and to obtain a $D$-dimensional regularized fermion propagator in Feynman diagrams.
The $D$-dimensional continuation of the interaction with the gauge boson admits several valid options, but we opt for the purely $4$-dimensional coupling as in \cite{Belusca-Maito:2020ala, Belusca-Maito:2021lnk,Martin:1999cc, Stockinger:2023ndm}. Refs.~\cite{Cornella:2022hkc,OlgosoRuiz:2024dzq} and \cite{Ebert:2024xpy} have studied successively wider sets of alternative options for definition of the $D$-dimensional Lagrangian. In the language of \cite{Ebert:2024xpy} we choose option 2b without evanescent gauge couplings and with purely right-handed $4$-dimensional pieces. This choice is motivated since it has been shown to lead to the simplest structure of the breaking of BRST invariance and thus the simplest form of the symmetry-restoring counterterms.

With these choices the fermion Lagrangian takes the familiar form
\begin{equation}\label{Eq:KineticTerms}
\begin{aligned}
    \mathcal{L}_{\mathrm{kin}+\mathrm{int}}^{\mathrm{fermion}}&=i\overline{\psi}_{i}\slashed{\partial}\psi_{i}+g T^a_{\mathcal{R}ij}\,\overline{\psi}_{Ri}\overline{\slashed{G}}^a\psi_{Rj}\\&=i\overline{\psi}_{Ri}\widehat{\slashed{\partial}}\psi_{Li}+i\overline{\psi}_{Li}\widehat{\slashed{\partial}}\psi_{Ri}+i\overline{\psi}_{Li}\overline{\slashed{\partial}}\psi_{Li}+\overline{\psi}_{Ri}\overline{\slashed{D}}_{\mathcal{R}ij}\psi_{Rj},
\end{aligned}
\end{equation}
where the purely 4-dimensional right-handed part has been written with the  covariant derivative
\begin{equation}
    D_{\mathcal{R}ij}^{\mu}=\partial^{\mu}\delta_{ij}-igT^a_{\mathcal{R}ij}G^{a\mu},
\end{equation}
whereas the additional terms in the second line result from the mismatch between $D$-dimensional derivative and $4$-dimensional gauge field in the first line.
The sterile left-handed fields have zero BRST transformations but transform under global gauge transformations such as to ensure charge conservation.

The full Lagrangian also contains the gauge part, ghosts, gauge-fixing and external sources (cf.~\cite{Belusca-Maito:2020ala}),
\begin{equation}\label{Eq:TheModel}
\mathcal{L}_{\mathrm{cl}}=\mathcal{L}_{\mathrm{kin}+\mathrm{int}}^{\mathrm{fermion}}+\mathcal{L}_{\mathrm{gauge}}+\mathcal{L}_{\mathrm{ghost}}+\mathcal{L}_{\mathrm{g-fix}}+\mathcal{L}_{\mathrm{ext}}.\end{equation}
The pure gauge boson part is given by
\begin{equation}
    \mathcal{L}_{\mathrm{gauge}}=-\frac{1}{4}F^2 \quad \mathrm{with} \quad F^{a\mu\nu}=\partial^{\mu}G^{a\nu}-\partial^{\nu}G^{a\mu}+g\epsilon^{abc}G^{b\mu}G^{c\nu},
\end{equation}
where $F^{a\mu\nu}F_{a\mu\nu}\equiv F^2$.
The third term in Eq.~(\ref{Eq:TheModel}) contains the covariant derivative of the ghost field in the adjoint representation and gives rise to the ghost-antighost kinetic term and the ghost-antighost interaction with the gauge boson,
\begin{equation}
    \mathcal{L}_{\mathrm{ghost}}=-\overline{c}\partial D_{\mathcal{A}}c=-\overline{c}_a\partial_{\mu}(\partial^{\mu}\delta^{ab}-g\varepsilon^{abc}G^{c\mu})c_b.
\end{equation}
The terms with the auxiliary Nakanishi-Lautrup field $B^a$ make up the gauge fixing part,
\begin{equation}
    \mathcal{L}_{\mathrm{g-fix}}=B(\partial G)+\frac{\xi B^2}{2},
\end{equation}
and could be simplified by the equations of motions $B^a=-\frac{1}{\xi}(\partial G^a)$.
The final term in Eq.~(\ref{Eq:TheModel}) contains the external sources $\{\rho^{a\mu},\zeta_a,\overline{R}^{i}_{\alpha},R^{j}_{\beta},\chi_a \}$ which are coupled to the (mostly) non-linear BRST transformations $s_D$ of the dynamical fields of the theory $\{G^{a\mu},c^a,\psi^{j}_{\beta},\overline{\psi}^{i}_{\alpha},\overline{c}^a\}$,
\begin{equation}\label{Eq:ExternalFields}
\mathcal{L}_{\text{ext}}=\rho^{\mu}_a s_D G^a_{\mu}+\zeta_a s_D c^a+\overline{R}^{i}s_D\psi_{Ri}+R^is_D\overline{\psi_{Ri}}+\chi_a s_D\overline{c}^a.\end{equation}
They allow us to express in a succinct and useful form the defining symmetry relations for 1-particle irreducible (1PI) Green functions and are classical fields which which do not propagate in loops. Their statistics, dimension and transformation properties (cf.~Tab.~\ref{Table:ExtSrc}) are defined to accommodate the respective quantum fields.
\begin{table}[!h]
    \begin{center}
    \begin{tabular}{|c||c|c|c|c|} \hline 
        External Field & Statistics & Dimension & Ghost Number & Lorentz Tr. \\ \hline \hline
        $\rho^{\mu}_a$ & Fermion & $3$ & $1$ & Four-Vector \\ \hline
        $\zeta^a$ & Boson & $4$ & $2$ & Scalar \\ \hline
        $\overline{R}^i_{\alpha}/R^{j}_{\beta}$ & Boson & $\frac{5}{2}$ & $1$ & Spinor\\ \hline
        $\chi^a$ & Boson & $2$ & $0$ & Scalar \\ \hline
    \end{tabular}
    \caption{External sources appearing  in the operators of Eq.~(\ref{Eq:ExternalFields}), and their properties. Together with the corresponding quantum fields the operators become bosonic Lorentz scalars of dimension $4$ and ghost number $0$.}
\label{Table:ExtSrc}
    \end{center}
    \end{table}
The  $D-$dimensional BRST transformations are defined as
\begin{equation}\label{Eq:sDDefinition}
    \begin{aligned}
        s_D(\psi_{Ri})&=igT_{\mathcal{R}ij}^ac^a\psi_{Rj},\\
        s_D(\overline{\psi}_{Ri})&=-igT_{\mathcal{R}ji}^ac^a\overline{\psi}_{Rj},\\
        s_D(\psi_{Li})&=0,\\
        s_D(G^{a\mu})&=\partial^{\mu}c^a+g\epsilon^{abc}G^{b\mu}c^c,\\
        s_D(c^a)&=-\frac{1}{2}g\epsilon^{abc}c^bc^c,\\
        s_D(\overline{c}^a)&=B^a,\\
        s_D(B^a)&=0.
    \end{aligned}
\end{equation}
The antighost $\overline{c}^a$ and the $B^a$-field form a BRST doublet and the sterile left-handed fermion as well as the external fields have vanishing BRST transformations. All other BRST transformations are non-linear in the dynamical fields.

By inspecting Eq.~(\ref{Eq:KineticTerms}) we find the usual BRST breaking (cf.~\cite{Belusca-Maito:2020ala, Belusca-Maito:2021lnk, Stockinger:2023ndm}) in $D$-dimension located in the evanescent kinetic term of the right-handed fermions,
\begin{equation}\label{Eq:TreeLevelLagrangianBreaking}
    \mathcal{S}_D\left(S_0\right) = 
     S_D\left(\Dintx \, \mathcal{L}_{\mathrm{cl}}\right)=
    \widehat{\Delta} \, \equiv 
     \Dintx \,
   gT_{\mathcal{R}ij}^a c^a
    \left[
    \overline{\psi}_i \left(\overset{\leftarrow}{\widehat{\slashed{\partial}}} \mathbb{P}_{\mathrm{R}} 
    + 
    \overset{\rightarrow}{\widehat{\slashed{\partial}}} \mathbb{P}_{\mathrm{L}} \right) \psi_j
    \right],
\end{equation}
which can be interpreted in terms of Feynman rules as the vertex,
\begin{equation}\label{Eq:TreeDeltaDef}
    \includegraphics[scale=.6, valign=c]{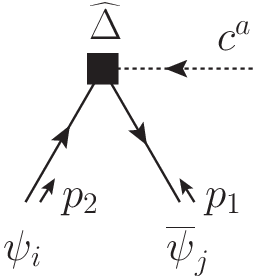} = gT^a_{\mathcal{R}ij}(\widehat{\slashed{p}_1}\projR+\widehat{\slashed{p}_2}\projL).
\end{equation}

\section{Determining the finite two-loop counterterms and non-Abelian obstacles}\label{Sec:DeterminationOfFCT}

This section serves as a brief discussion of our approach to determining the spurious symmetry breaking and restoring the symmetry via counterterms. For a thorough account see Ref.\ \cite{Belusca-Maito:2023wah}. The main focus will be to highlight differences stemming from non-Abelian interactions. The structure constants are no longer zero, which e.g. makes for interaction terms between ghosts and gauge bosons. Although this has been the case for related works in non-Abelian models at one-loop as well, some of these complications appear first at the two-loop level, while others are technical issues due to the necessity of implementing the external source operators in computer calculations. We shall begin with a brief exposition of Slavnov-Taylor identities and the quantum action principle which are our means of quantifying and computing the symmetry breaking while also allowing us to shed light on some non-Abelian aspects. In the second part we explain how we implemented both the breaking $\Delta$-vertices and external fields in \texttt{FeynArts}.

\subsection{Slavnov-Taylor identities and the quantum action principle in DReg}\label{Sec:STIProcedure}

The appropriate tool for studying BRST symmetry in the renormalized theory is the Slavnov-Taylor identity, which expresses BRST invariance on the level of 1PI Green functions. The ultimate Slavnov-Taylor identity required to hold after renormalization is
\begin{equation}\label{Eq:STIRenCond}
    \underset{D\rightarrow 4}{\mathrm{LIM}}\,\mathcal{S}_D(\Gamma_{\mathrm{DRen}})=0,
\end{equation}
where the limit includes setting evanescent objects to zero. 
Here the Slavnov-Taylor operator is defined as
\begin{equation}\label{Eq:STIRenNew}
    \mathcal{S}(\Gamma_{\mathrm{DRen}})=\intx\,\frac{\delta \Gamma_{\mathrm{DRen}}}{\delta K_{\phi}}\frac{\delta \Gamma_{\mathrm{DRen}}}{\delta\phi}
    ,
\end{equation}
where we follow the notation of Ref.\ \cite{Belusca-Maito:2023wah} for the 1PI quantum effective action $\Gamma$, and where $\phi$ and $K_{\phi}$ denote the quantum fields and the sources introduced in Eq.~(\ref{Eq:ExternalFields}), respectively. We further introduce a linearized Slavnov-Taylor operator, 
\begin{equation}\label{Eq:bDDef}
b_D=s_D+\int d^Dx\,\frac{\delta S_0^{(D)}}{\delta G^{a\mu}}\frac{\delta}{\delta\rho^{a\mu}}+\frac{\delta S_0^{(D)}}{\delta \psi_{j\beta}}\frac{\delta}{\delta \overline{R}_{j\beta}}+\frac{\delta S_0^{(D)}}{\delta \overline{\psi}_{i\alpha}}\frac{\delta}{\delta R_{i\alpha}}+\frac{\delta S_0^{(D)}}{\delta c^a}\frac{\delta}{\delta\zeta^a}+\frac{\delta S_0^{(D)}}{\delta \overline{c}^a}\frac{\delta}{\delta\chi^a},
\end{equation}
where we denote the dimensionally regularized tree-level action $S_0^{(D)}\equiv S_0$, and $s_D$ is the BRST operator in $D$ dimensions whose action is defined in Eq.~(\ref{Eq:sDDefinition}), cf.~\cite{Belusca-Maito:2020ala}. The remaining terms in Eq.~(\ref{Eq:bDDef}) are given by derivatives with respect to external fields multiplied by equation of motion terms of the associated dynamical fields. The operator $b_D$ plays an important role in the symmetry restoration process. Its action captures the linearized impact of adding counterterms on the STIs. Its $4$-dimensional counterpart $b_4$ is nilpotent, $b_4^2=0$, and its cohomology governs potential symmetry breakings in the space of ghost number $=1$ and dimension $\leq4$ operators (cf.~\cite{Piguet:1995er} and also \cite{Belusca-Maito:2023wah}). Trivial elements in the cohomology are $b_4$-exact terms which correspond to spurious breakings that can be cancelled by local finite counterterms while non-exact closed forms amount to gauge anomalies, which are assumed absent here. 

Finally, thanks to the quantum action principle of DReg \cite{Breitenlohner:1975hg, Breitenlohner:1976te, Breitenlohner:1977hr, Belusca-Maito:2023wah}, it is possible to express a potential violation of Eq.~(\ref{Eq:STIRenCond}), i.e.~a non-vanishing value of Eq.~(\ref{Eq:STIRenNew}), via a symmetry-breaking operator insertion,
\begin{equation}\label{Eq:QAPDReg}
    \mathcal{S}_D(\Gamma_{\mathrm{DReg}})=\Delta\cdot\Gamma_{\mathrm{DReg}},
\end{equation}
where we generically denote $\Gamma_{\mathrm{DReg}}$ as the dimensionally regularized but not necessarily (sub-)renormalized effective action. The breaking insertion $\Delta$ can be computed order-by-order by applying the Slavnov-Taylor operator onto the action,
\begin{equation}\label{Eq:GeneralDefDelta}
    \Delta=\mathcal{S}_D(S_0+S_{\mathrm{ct}}),
\end{equation}
where $S_{\mathrm{ct}}=S_{\mathrm{sct}}+S_{\mathrm{fct}}$ denotes the ($D$-dimensional) counterterm action comprised of both divergent UV counterterms and finite symmetry-restoring ones. The equality (\ref{Eq:QAPDReg}) holds true with or without counterterms, provided the counterterms are included consistently on the LHS and the RHS.

Explicitly at two-loop order we can expand the RHS of  Eq.~(\ref{Eq:GeneralDefDelta}),
\begin{equation}\label{Eq:DeltaAt2Loop}
    \Delta^{\leq2\mathrm{L}}=\widehat{\Delta}+\Delta^{1\mathrm{L}}_{\mathrm{ct}}+\mathcal{S}_D(S_{\mathrm{ct}}^{1\mathrm{L}})+\Delta^{2\mathrm{L}}_{\mathrm{ct}},
\end{equation}
and the RHS of Eq.~(\ref{Eq:QAPDReg}),
\begin{equation}\label{Eq:DeltaGammaAt2Loop}
    \Delta\cdot\Gamma_{\text{DReg}}^{\leq2\mathrm{L}}
    =
    \widehat{\Delta}\cdot\Gamma_{\text{DReg}}^{\leq2\mathrm{L}}
    +\Delta^{1\mathrm{L}}_{\mathrm{ct}}\cdot\Gamma_{\text{DReg}}^{\leq2\mathrm{L}}
    +\mathcal{S}_D(S_{\mathrm{ct}}^{1\mathrm{L}})
    +
    \Delta^{2\mathrm{L}}_{\mathrm{ct}},
\end{equation}
with $\Delta^{n\mathrm{L}}_{\mathrm{ct}}=b_D(S_{\mathrm{ct}}^{n\mathrm{L}})$.
Note that the classical equations of motion of both $\chi^a$ and $B^a$ are at most linear and hence receive no loop corrections. The action of the derivatives with respect to these fields in $b_D$ on the counterterm action is therefore trivial.

The purely evanescent breaking at tree-level $\widehat{\Delta}$ is the one found in Eq.~(\ref{Eq:TreeLevelLagrangianBreaking}), while the updated breakings $\Delta^{1\mathrm{L}}_{\mathrm{ct}}$ contain $4$-dimensional and evanescent divergent parts as well as finite pieces. The first two terms in Eq.~(\ref{Eq:DeltaAt2Loop}) enter in Eq.~(\ref{Eq:DeltaGammaAt2Loop}) and require the calculation of actual two- and one-loop diagrams, respectively. 
The second to last term in Eq.~(\ref{Eq:DeltaAt2Loop}) is of two-loop order due to the non-linearity of $\mathcal{S}_D$ (cf. Eq.~(\ref{Eq:STIRenNew})),
and amounts to products of one-loop counterterm insertions.
Finally, the genuine two-loop counterterms needed for renormalization are given by the last term 
in Eq.~(\ref{Eq:DeltaAt2Loop}),
\begin{equation}
    \Delta^{2\mathrm{L}}_{\mathrm{ct}}=\Delta^{2\mathrm{L}}_{\mathrm{sct}}+\Delta^{2\mathrm{L}}_{\mathrm{fct}}=b_D(S^{2\mathrm{L}}_\mathrm{sct})+
b_D(S^{2\mathrm{L}}_\mathrm{fct}).
\end{equation}
Hence, after the theory has been one-loop renormalized, the procedure to determine the two-loop counterterms is as follows. First, the divergent two-loop counterterms  $S^{2\mathrm{L}}_{\mathrm{sct}}$  are defined such that all power-counting divergent two-loop Green functions become finite. The only remaining unknowns are then the finite two-loop counterterms $S^{2\mathrm{L}}_{\mathrm{fct}}$, which enter Eq.~(\ref{Eq:DeltaGammaAt2Loop}) via $\Delta^{2\mathrm{L}}_{\mathrm{fct}}$. They are determined by the requirement that Eq.~(\ref{Eq:DeltaGammaAt2Loop}) must vanish at the two-loop level in the limit $\mathrm{LIM}_{D\rightarrow 4}$, equivalently to the validity of Eq.~(\ref{Eq:STIRenCond}) at the two-loop level.
Since finite BRST-invariant counterterms do not contribute to Eq.~(\ref{Eq:STIRenCond}), the finite counterterms are not unique. 

The method sketched here was explained in detail generically in Refs.~\cite{Belusca-Maito:2020ala,Belusca-Maito:2023wah} and specified to the two-loop level in Ref.~\cite{Belusca-Maito:2021lnk}. Ref.~\cite{Cornella:2022hkc} followed a similar approach at the one-loop level, however based on the background field method and partially independent of DReg. An important advantage we shall employ throughout our analysis is that 
only UV power-counting divergent Green functions in Eq.~(\ref{Eq:DeltaGammaAt2Loop}) can contribute to the finite symmetry breaking via local, $4$-dimensional terms, while evanescent finite terms vanish in the limit $\mathrm{LIM}_{D\rightarrow 4}$ (cf.~Eq.~(\ref{Eq:STIRenCond})). Power-counting finite breaking Green functions may have evanescent finite terms, which likewise vanish, and hence can be neglected entirely.
The fact that $\Delta^{2\mathrm{L}}_{\mathrm{ct}}$ must be a $b_D$ variation of a finite, local expression amounts to very strong constraints on the possible results of the Green functions $\Delta \cdot \Gamma$ entering Eq.~(\ref{Eq:DeltaGammaAt2Loop}) and thus allows very nontrivial checks.

\subsection{Remarks on non-Abelian Subtleties}\label{SubSec:NASubtleties}
In this subsection we comment in some detail on ways in which the two-loop calculations become more complicated, compared to the calculations in the Abelian case of Refs.~\cite{Belusca-Maito:2021lnk,Stockinger:2023ndm}. This includes remarks on non-Abelian subtleties in
the structure of the STIs, the increased number of Green functions as well as explicit checks of Eq.~(\ref{Eq:QAPDReg}).

\paragraph{non-Abelian structures of the breaking}
~\\

In Abelian models, ghosts do not interact and operators containing external sources do not receive perturbative corrections. Hence the Abelian counterterm action does not contain external sources at any order \cite{Belusca-Maito:2021lnk}. When evaluating the breaking insertion (\ref{Eq:GeneralDefDelta}) in the Abelian case, the operator $b_D$ simply reduces to $s_D$ in Eq.~(\ref{Eq:DeltaAt2Loop}), and the second to last term of Eq.~(\ref{Eq:DeltaAt2Loop}) vanishes. All of this changes in the non-Abelian case. Exemplarily, Refs.~\cite{Martin:1999cc, Belusca-Maito:2020ala} obtained a 1-loop counterterm of the $\overline{R}/R$-field with divergent and symmetry-restoring finite contributions,
\begin{align}\label{Eq:exemplifysourcect}
    S_{\mathrm{ct}}^{1\mathrm{L}} &\supset -\frac{g^2\xi C_2(G)}{16\pi^2}\left(\frac{1}{2\epsilon}+\frac{1}{4}\right)\int d^Dx\,\overline{R} c \psi_{R} + h.c..
\end{align}

As a first noteworthy consequence, in the evaluation of  $\Delta^{1\mathrm{L}}_{\mathrm{ct}}=b_D(S_{\mathrm{ct}}^{1\mathrm{L}})$ in Eq.~(\ref{Eq:DeltaAt2Loop}), the action of $b_D$  onto just the finite part $S_{\mathrm{fct}}^{1\mathrm{L}}$ produces finite and evanescent contributions of the form
\begin{equation}\label{Eq:Delta1FCTEvanescentContribution}
   \Delta^{1\mathrm{L}}_{\mathrm{fct}} \supset
   \int d^{D}x \,\frac{\delta S_0}{\delta \psi_{i\alpha}}\frac{\delta S^{1\mathrm{L}}_{\mathrm{ct}}}{\delta\overline{R}_{i\alpha}}
   \sim \overline{\psi} i(\overline{\slashed{\partial}}+\widehat{\slashed{\partial}})c\psi,
\end{equation}
when the $D$-dimensional fermion kinetic term within $S_0$ and the term given in Eq.~(\ref{Eq:exemplifysourcect}) are considered. The appearance of the evanescent operator $\widehat{\slashed{\partial}}$ within $\Delta^{1\mathrm{L}}_{\mathrm{fct}}$ is a new non-Abelian feature. Taking such terms into account is crucial for a consistent renormalization at higher loop orders. 

Likewise the second to last term of Eq.~(\ref{Eq:DeltaAt2Loop}) is of a purely non-Abelian nature. Working it out by just inserting the term given in Eq.~(\ref{Eq:exemplifysourcect}) yields
\begin{equation}\label{Eq:STIofSct1New}
\mathcal{S}_D(S_{\mathrm{ct}}^{1\mathrm{L}})
\supset
\int d^{D}x \,\frac{\delta S^{1\mathrm{L}}_{\mathrm{ct}}}{\delta \psi_{i\alpha}}\frac{\delta S^{1\mathrm{L}}_{\mathrm{ct}}}{\delta\overline{R}_{i\alpha}}
   \sim
   \overline{R} c c \psi,
\end{equation}
which corresponds to finite as well as UV divergent contributions to the breaking Green function $(\Delta\cdot\Gamma)_{\psi\overline{R} c c}$. Including such terms is therefore vital for the UV finiteness and overall correctness of breaking Green functions.

\paragraph{Computational complexity}
~\\

Apart from the obvious growth of the number of diagrams, due to gauge boson self-interactions and ghost interactions there are significant changes to the structure of possible diagrams and to the set of relevant Green functions, compared both to the Abelian case \cite{Belusca-Maito:2021lnk,Stockinger:2023ndm} and to the non-Abelian one-loop case \cite{Martin:1999cc,Belusca-Maito:2020ala}. 

First, as mentioned above, in the Abelian case there are no loop corrections to Green functions with external sources. In the non-Abelian case there are. E.g.~the one-loop counterterm mentioned in Eq.~(\ref{Eq:exemplifysourcect}) corresponds to the breaking Green function
\begin{align}
    (\Delta\cdot\Gamma)_{\psi\overline{R} c c},
\end{align}
which is non-vanishing at the one-loop order. At the two-loop level, there are additional non-vanishing Green functions with external sources contributing to the breaking Eq.~(\ref{Eq:DeltaGammaAt2Loop}), namely 
\begin{align}\label{Eq:ExamplesOfBeakingSourcesAt2Loop}
    &(\Delta\cdot\Gamma)_{\rho c c},\nonumber\\
    &(\Delta\cdot\Gamma)_{G c c\rho},\\
    &(\Delta\cdot\Gamma)_{\zeta c c c}\nonumber.
\end{align}
The genuine two-loop diagrams for these Green functions involve an insertion of the tree-level breaking vertex $\widehat{\Delta}$ and a fermion loop, which then couples to the external sources via a second loop.

The same is true for Green functions whose external lines are made up of gauge bosons or ghosts exclusively, such as
\begin{align}
    (\Delta\cdot\Gamma)_{\overline{c}cc},
\end{align}
which appears for the first time at the two-loop level.

Another noteworthy consequence of interacting ghosts is that each line of the $\Delta$-vertex insertion can be fully internal to the diagram. As a consequence while at one-loop there were only up to 4-point functions with non-vanishing contributions, now at two-loop even 5-point functions become relevant. For example the computationally most costly Green function of the entire two-loop calculation  is
\begin{align}
    (\Delta\cdot\Gamma)_{cGGGG},
\end{align}
a five-point function with five open colour indices.
This Green function must be computed at the two-loop level and may provide a non-vanishing contribution. Indeed it is found to contribute to the divergent part of the symmetry breaking in our model.
In contrast, in the Abelian case such a Green function cannot receive contributions either divergent or finite and non-evanescent
at any order because any local breaking could not be written as the BRST variation of a dimension $\leq 4$-counterterm. Also, in the the non-Abelian one-loop calculations of Refs.~\cite{Martin:1999cc,Belusca-Maito:2020ala} this Green function was irrelevant since it became effectively power-counting finite due to the particular structure of the $\widehat{\Delta}$-vertex Feynman rule in Eq.~(\ref{Eq:TreeDeltaDef}), which can be seen to cancel the highest power in loop momenta if the ghost line of the $\widehat{\Delta}$-vertex is external to the diagram. These finite terms vanish in Eq.~(\ref{Eq:STIRenCond}), as announced at the end of Sec.~\ref{Sec:STIProcedure}.

\paragraph{Checking the quantum action principle}
~\\

In the Abelian two-loop study of Ref.~\cite{Belusca-Maito:2021lnk} we verified the correctness of our procedure by checking the validity of certain Ward identities dictated by an intact Slavnov-Taylor identity. The simplest example is given by the transversality of the gauge boson self-energy,
\begin{equation}\label{Eq:TransversalityPhotonSelfEnergyRenormalized}
    p_{\nu}\Gamma^{\mathrm{Ren}}_{G^{a\nu}G^{b\mu}(-p)}=0,
\end{equation}
which was valid once the appropriate symmetry-restoring counterterms had been found.
As explained in Sec.~\ref{Sec:STIProcedure}, our method is to determine the counterterms via Eq.~(\ref{Eq:DeltaGammaAt2Loop}), corresponding to the RHS of the quantum action principle (\ref{Eq:QAPDReg}). The check then corresponds to an explicit evaluation of the LHS of Eq.~(\ref{Eq:QAPDReg}), corresponding to the Ward and Slavnov-Taylor identities. Indeed for the subrenormalized action Eq.~(\ref{Eq:QAPDReg}) yields in this sector,
\begin{equation}\label{Eq:TransversalityPhotonSelfEnergySubRenormalized}
    \Gamma^{\mathrm{DReg}}_{c^a\rho^{c\nu}}\Gamma^{\mathrm{DReg}}_{G^{c\nu}G^{b\mu}(-p)}=(\Delta\cdot\Gamma)^{\mathrm{DReg}}_{G^{b\mu}c^a(p)}.
\end{equation}
The equation must be valid by virtue of the quantum action principle. However, explicitly checking it is highly nontrivial since the structures on the RHS, where we exploit (evanescent) operator insertions and only need to evaluate UV poles, are manifestly different from the ones on the LHS, where ordinary Green functions appear whose finite parts matter.

Thus it is desirable to also perform such quantum action principle checks in the non-Abelian case in as many sectors as possible. While in the Abelian case the important Ward identities correspond to a few rather straightforward relations between ordinary Green functions, the non-Abelian identites are more involved. Only Eq.~(\ref{Eq:TransversalityPhotonSelfEnergyRenormalized}) manages to preserve its simple structure for the finite breaking in the non-Abelian case, albeit still subject to more numerous contributions in the full intermediate steps of the LHS of Eq.~(\ref{Eq:TransversalityPhotonSelfEnergySubRenormalized}).

To illustrate the nature of these checks and the appearing difficulties we focus on the example relation
\begin{equation}
\begin{aligned}\label{Eq:ExampleQAP3Pt}
    \Gamma^{\mathrm{DReg}}_{c^a\rho^{b\mu}}\Gamma^{\mathrm{DReg}}_{\psi_{j\beta}\overline{\psi}_{i\alpha} G^{b\mu}}+
    \Gamma^{\mathrm{DReg}}_{\psi_{j\beta}c^a\overline{R}_{k\delta}}\Gamma^{\mathrm{DReg}}_{\psi_{k\delta}(-p_1)\overline{\psi}_{i\alpha}(p_1)}&-
    \Gamma^{\mathrm{DReg}}_{R_{k\delta}c^a\overline{\psi}_{i\alpha}}\Gamma^{\mathrm{DReg}}_{\psi_{j\beta}(p_2)\overline{\psi}_{k\delta}(-p_2)}\\
    &=(\Delta\cdot\Gamma^{\mathrm{DReg}})_{\psi_{j\beta}(p_2)\overline{\psi}_{i\alpha}(p_1)c^a(p)},
    \end{aligned}
\end{equation}
which is obtained by functional differentiation of both the LHS and RHS of Eq.~(\ref{Eq:QAPDReg}) w.r.t.\ the fields $\{c^a,\overline{\psi}_{i\alpha},\psi_{j\beta}\}$. On the right-hand side we find the breaking Green function $(\Delta\cdot\Gamma)_{\psi\overline{\psi}c}$ just as in Ref.~\cite{Belusca-Maito:2021lnk}. This Green function is computed as part of the procedure to determine the symmetry-restoring counterterms both in the Abelian and the non-Abelian case. The LHS conversely yields sums of products of ordinary Green functions.
The equality between the LHS and RHS must hold both for the divergent and the finite parts. If the effective action is fully renormalized up to two-loop order, they need to evaluate to zero. If the action is merely subrenormalized, the products of Green functions on the LHS need to give the same poles and finite breakings as the subrenormalized insertion of $\widehat{\Delta}$ on the RHS. We perform the checks for Eq.~(\ref{Eq:TransversalityPhotonSelfEnergySubRenormalized}) and Eq.~(\ref{Eq:ExampleQAP3Pt}) at the level of the subrenormalized effective action, as in Ref.~\cite{Belusca-Maito:2021lnk}.

\begin{figure}
\centering
\begin{tabular}{c@{\hskip 0.75in}c@{\hskip 0.75in}c@{\hskip 0.75in}}
\(\begin{gathered}
    \includegraphics[scale=.45]{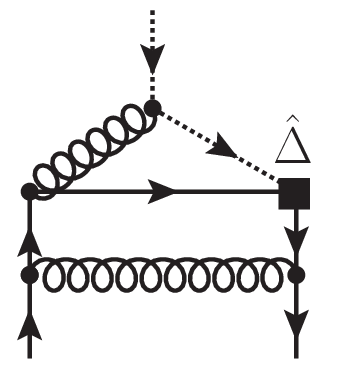}
    \end{gathered}\)
    &
    \(
    \begin{gathered}
    \includegraphics[scale=.45]{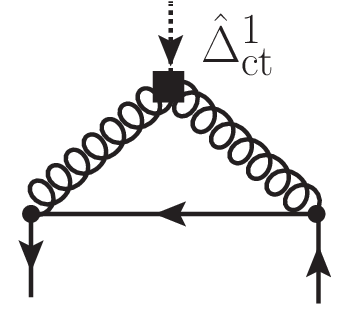}
    \end{gathered}\)
    &
    \(
\begin{gathered}
    \includegraphics[scale=.45, valign=c]{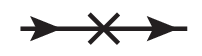}\hspace{1ex}\times \includegraphics[scale=.45, valign=c]{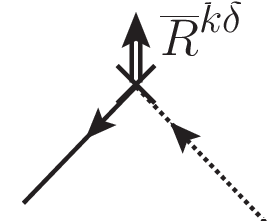}
    \end{gathered}
    \)
    \\
\end{tabular}
\caption{Examples of three kinds of contributions which appear on the RHS of Eq.~(\ref{Eq:ExampleQAP3Pt}). The first is a pure two-loop breaking diagram with fully internal $\widehat{\Delta}$-vertex. The middle exemplifies  one-loop diagrams with one-loop $\Delta^1_{\mathrm{ct}}$-counterterm insertions. The third example shows a product of pure counterterms due to the second to last term in Eq.~(\ref{Eq:DeltaAt2Loop}).}
\label{Fig:ExampleDiagramsDeltaCPP2LRHS}
\end{figure}

In the Abelian two-loop calculation the LHS of Eq.~(\ref{Eq:ExampleQAP3Pt}) simplifies dramatically and produces the well-known QED-like Ward identity between the fermion self-energy and the fermion-gauge boson vertex correction. Indeed in the Abelian case the first factor $\Gamma^{\mathrm{DReg}}_{c^a\rho^{b\mu}}$ simply becomes the photon momentum, cf.~Eq.~(\ref{Eq:ExternalFields}), since the  Green functions with external sources receive no higher order corrections. The remaining two terms simplify in an analogous way. The fermion source operators take on their tree-level expressions, and the two fermion self energies hence come with similar coefficients, albeit different external momenta,
\begin{equation}
    p^{\mu}\Gamma^{\mathrm{Ren}}_{\psi\overline{\psi}G}(p_2,p_1,p)+eQ\left(\Gamma^{\mathrm{Ren}}_{\psi\overline{\psi}}(p_2,p+p_1)-\Gamma^{\mathrm{Ren}}_{\psi\overline{\psi}}(p+p_2,p_1)\right)=0.
\end{equation}
In the non-Abelian case, both the LHS and the RHS of Eq.~(\ref{Eq:ExampleQAP3Pt}) lead to additional complications. First we discuss the kinds of contributions appearing on the RHS of Eq.~(\ref{Eq:ExampleQAP3Pt}).  Fig.~\ref{Fig:ExampleDiagramsDeltaCPP2LRHS} illustrates three classes of diagrams.

The first diagram is purely two-loop and carries an internal insertion of the tree-level $\widehat{\Delta}$-vertex. As discussed this is a markedly non-Abelian example. There are of course further pure two-loop diagrams. Some of them exist in the Abelian model and are either structurally identical or outwardly Abelian but with internal non-Abelian corrections. Then there are non-Abelian types of diagrams which were present at one loop (cf.~Ref.~\cite{Belusca-Maito:2020ala})
but now with an additional internal correction and for which the $\widehat{\Delta}$-vertex may still be external but its ghost line is internal.
The second class of diagrams comprises the counterterm corrections through $b_D(S^{1\mathrm{L}}_{\mathrm{ct}})$, which are largely similar to the previous papers \cite{Belusca-Maito:2021lnk, Stockinger:2023ndm}. Still, for the first time we get a contribution in $\Delta^{1\mathrm{L}}_{\mathrm{fct}}$ which contains both $4$-dimensional as well as evanescent (cf.~Eq.~(\ref{Eq:Delta1FCTEvanescentContribution})) terms. Finally we see an instance of the second to last term in Eq.~(\ref{Eq:DeltaAt2Loop}). Here the product of one-loop divergent and finite counterterms gives rise to both divergent and finite contributions and is hence indispensable for the correct finite counterterms.

\begin{figure}
\centering
 \begin{tabular}{c c c}
\(\displaystyle \delta^{ab}p^{\mu}\,\times\hspace{-1.3ex}\begin{gathered}\includegraphics[scale=.41]{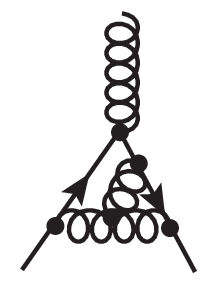}\end{gathered}\quad-igT^b_{\mathcal{R}ij}(\overline{\gamma}^{\mu}\mathbb{P}_R)_{\alpha\beta}
\,\times\hspace{-1.3ex}\begin{gathered}\includegraphics[scale=.41]{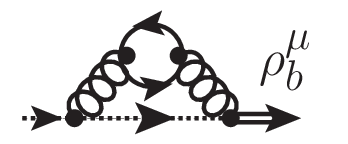}\end{gathered}\)& \(\displaystyle\) &
\(\displaystyle\begin{gathered}\includegraphics[scale=.41]{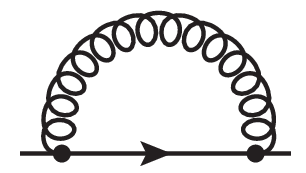}\end{gathered}\times\hspace{-1ex}\begin{gathered}\includegraphics[scale=.45]{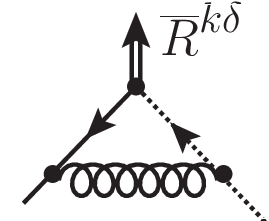}\end{gathered}\)
\end{tabular}
\caption{The first two combinations correspond to $(\text{tree-level})\times(\text{two-loop})$ and $(\text{two-loop})\times(\text{tree-level})$ for the Green functions $\Gamma_{c^b(p)\rho^{b\mu}}$ and $\Gamma_{G^{b\mu}\psi_{j\beta}\overline{\psi}_{i\alpha}}$. The third combination corresponds to $(\text{one-loop})\times(\text{one-loop})$ for the Green functions $\Gamma_{\psi_{k\delta}\overline{\psi}_{i\alpha}(p_1)}$ and $\Gamma_{\psi_{j\beta}c^a\overline{R}_{k\delta}}$.
}
\label{Fig:ExampleDiagramsDeltaCPP2LLHS}
\end{figure}

Turning to the LHS of Eq.~(\ref{Eq:ExampleQAP3Pt}) each 1PI Green function contains tree-level, one-loop and two-loop contributions and so the product contains terms ranging from tree-level to four-loop order. In the non-Abelian case, all these combinations are typically non-zero, and the appropriate combinations need to be selected in order to evaluate the identity at the two-loop level. The first two combinations shown in Fig.~\ref{Fig:ExampleDiagramsDeltaCPP2LLHS} correspond to the first term in Eq.~(\ref{Eq:ExampleQAP3Pt}), where either Green function is evaluated at tree-level or two-loop level. The first such term involves an ordinary two-loop vertex correction and has Abelian counterparts. The second term however only arises in the non-Abelian case and corresponds to higher order corrections to the external source operators, here to $\rho^{b\mu}$, corresponding to the BRST transformation of the gauge field. 
The third combination of Fig.~\ref{Fig:ExampleDiagramsDeltaCPP2LLHS} comes from the one loop diagrams generated by the products of effective action terms for the fermions and fermion sources. It is in general not sufficient in this case to evaluate merely the finite part, let alone the poles, but even terms up to $\mathcal{\epsilon}$ have to be kept as they become finite terms upon hitting poles of the other Green function.
They are likewise ruled out in the Abelian calculation.

The non-Abelian differences we have discussed here by means of the example of Eq.~(\ref{Eq:ExampleQAP3Pt}) illustrate the manifold ingredients of the full two-loop renormalization and exemplify the higher complexity of quantum action principle checks. Still, we have carried out such consistency checks in all sectors, i.e.~for all breaking Green functions. The check for the specific cases of Eq.~(\ref{Eq:TransversalityPhotonSelfEnergyRenormalized}) and Eq.~(\ref{Eq:ExampleQAP3Pt}) has been done including the finite parts, the checks in all other sectors on the level of UV poles. The number of these checks is thus much higher than in the Abelian case, and include features such as two-loop diagrams of external sources or products of one-loop counterterm contributions for all sectors. 

\subsection{Implementation in \texttt{FeynArts}}
For any systematic multi-loop calculation it is crucial that the computation takes place in a unified, automated setup. A particular complication in the present case stems from exotic objects like the fermionic, ghost-number-one $\Delta$ vertex and the BRST operators coupled to external sources, as well as their interplay. In Abelian calculations these issues presented no serious obstructions. In the non-Abelian one-loop calculation of Ref.\ \cite{Belusca-Maito:2020ala} some of these subtleties could be circumvented since for example the contributions of external sources amounted to but a few diagrams and could be handled manually, which is no longer feasible at two-loop. It is therefore desirable to create a working implementation which addresses these specialties, and is capable of automatically handling the two-loop calculation while allowing for suitable generalizations. Such a successful implementation around a  \texttt{FeynArts}-model file will be discussed here briefly.

\begin{figure}
\centering
\begin{center}
\begin{tabular}{c c c c c c c}
\includegraphics[scale=.45]{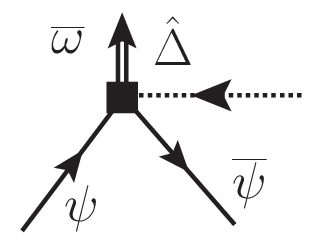}&
\raisebox{22pt}{$\Longrightarrow$}
&
\includegraphics[scale=.45]{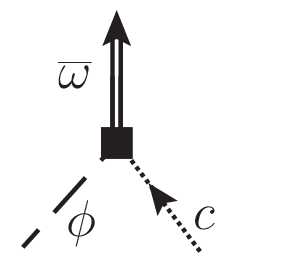}&
\raisebox{22pt}{\(\!\!\!\!\otimes\)} &
\includegraphics[scale=.45]{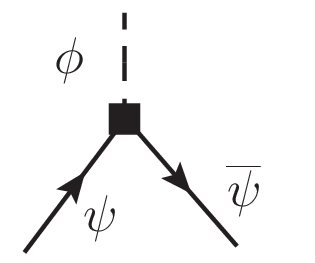}&
\raisebox{22pt}{$\Longrightarrow$}
&
\includegraphics[scale=.45]{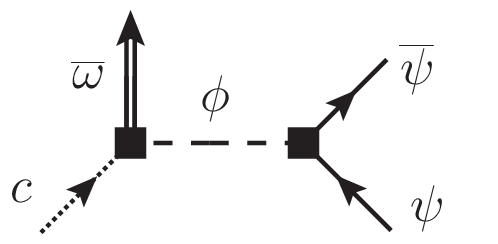}
\end{tabular}
\end{center}
\caption{Schematic decomposition of the $\widehat{\Delta}$-vertex coupled to an antighost source $\overline{\omega}$ into two normal three-point vertices $\phi\overline{c}c$ and $\phi\overline{\psi}\psi$.
These form a reducible tree-level diagram connected by an auxiliary scalar field $\phi$ with unit propagator to reproduce the correct Feynman rule of $\widehat{\Delta}$.
}
\label{fig:DeltaVertexDecomposition}
\end{figure}

As a fermionic object with ghost number $-1$, the tree-level $\widehat{\Delta}$-operator is very different from a typical Lagrangian term, yet is effectively treated as such in the logic of \texttt{FeynArts}. In the quantum action principle (cf.~Eq.~(\ref{Eq:QAPDReg})) $\widehat{\Delta}$ is understood as a one-time insertion into the action which can be defined by introducing an additional external source which couples to the operator and renders it a standard Lagrangian term,
\begin{equation}
    \widehat{\Delta}\cdot\Gamma_{\phi_1\dots\phi_n}=\frac{\delta\Gamma_{\phi_1\dots\phi_n}}{\delta \overline{\omega}_{\Delta}}\bigg\vert_{\overline{\omega}_{\Delta}=0},
\end{equation}
where $\overline{\omega}_{\Delta}$ denotes an auxiliary antighost. Hence it would be suitable to realize the breaking vertex as $\overline{\omega}_{\Delta}c\overline{\psi}\psi$. Since however in \texttt{FeynArts} handling of such $4$-fermion vertices can be subtle, it is preferable to split them into two $3$-point vertices by means of an auxiliary scalar field $\phi$ with unit propagator, as shown in Fig.~\ref{fig:DeltaVertexDecomposition}.
On the Lagrangian level this can be schematically understood via eliminating an auxiliary scalar field by its equations of motion,
\begin{equation}\label{Eq:MotivateFourPointDeltaVertexByEoms}
    \mathcal{L}_{\phi-\mathrm{aux}}=
    -\frac{\phi^2}{2}+\overline{\omega}_{\Delta}c^a\phi^a+gT^a_{\mathcal{R}ij}\left(\overline{\psi}_i\overset{\leftarrow}{\widehat{\slashed{\partial}}}\projR\psi_j+\overline{\psi}_i\overset{\rightarrow}{\widehat{\slashed{\partial}}}\projL\psi_j\right)\phi^{a}.
\end{equation}
The Feynman rules of the $3$-point vertex insertions $\overline{\omega}_{\Delta}c\phi$ and $\overline{\psi}\psi\phi$ are adjusted in such a way that the reducible tree-level diagram rightmost in Fig.~\ref{fig:DeltaVertexDecomposition} and formed from these vertices reproduces the familiar Feynman rule of Eq.~(\ref{Eq:TreeDeltaDef}).

For multi-loop calculations of $\widehat{\Delta}$-Green functions thus implemented there will be both 1PI reducible and irreducible diagrams, the former coming from diagrams where the original ghost $c^a$ within $\widehat
{\Delta}$ and thus the ghost of the left vertex in the middle of Fig.~\Ref{fig:DeltaVertexDecomposition} is external to the diagram. These include all diagrams of the Abelian model. The diagram generation has to be furnished with additional selection rules such that only sensible reducible diagrams are considered.

The external sources for BRST transformations are external fields with antighost number and wrong statistics; hence their implementation in \texttt{FeynArts} employs similar decompositions. At first, the external sources  are effectively treated as composite fields of conventional types available in \texttt{FeynArts}, e.g.
\begin{equation}\label{Eq:DefinitionFeynArtsSources}
\rho^{a\mu}\longrightarrow\overline{\omega}_{\rho}^aA^{\mu}\qquad R_{j\beta}\longrightarrow\overline{\omega}_{R}\chi_{j\beta}.
\end{equation}
Here $A^{\mu}$ denotes an auxiliary vector field and $\chi_{j\beta}$ refers to a fermion, and $\overline{\omega}_{\rho,R}$ denote further auxiliary antighosts. For instance the $\overline{\psi}cR$ vertex becomes the 4-point vertex $\overline{\psi}c \overline{\omega}_{R}\chi$. In a second step in this and all similar cases, 4-point interactions of fermionic fields are treated analogously to $\widehat{\Delta}$ and split into two $3$-point vertices. This also means that an $n$-point Green function with insertions of either $\Delta$ or external sources is represented in \texttt{FeynArts} as a higher-valence function.

We couple each breaking vertex, i.e.~also those generated at one-loop in $\Delta^{1\mathrm{L}}_{\mathrm{ct}}$, by different external ghosts and scalars and can access them individually for selecting diagrams in \texttt{FeynArts}. The same decomposition as in Fig.~\ref{fig:DeltaVertexDecomposition} is used if needed.
The most delicate kind of higher-valence vertices in our implementation
due to the one-loop renormalization is given by the breaking operator $\Delta^{1\mathrm{L}}_{\psi c c\overline{R}}$. Since it is both a $\Delta$-insertion and carries the external source $\overline{R}$, this operator necessitates two additional antighost sources, which would in total correspond to a vertex of valence six. With two spinors, two ghosts and two antighosts, the vertex is decomposed into three three-point vertices each with their own scalar field, and joined by a triple scalar vertex. Despite this complication, we recover the necessary diagrams for $(\Delta^{1\mathrm{L}}_{\psi cc\overline{R}}\cdot\Gamma)^{2\mathrm{L}}$.

Our implementation as described here correctly reproduces the results in Refs.~\cite{Belusca-Maito:2020ala,Martin:1999cc}. We mention that simpler implementations in \texttt{FeynArts} which e.g.~do not couple the $\widehat{\Delta}$-vertex to an auxiliary source field have been tested but produce incorrect results.
Explicitly, at one-loop the Green function $(\widehat{\Delta}\cdot\Gamma)_{\psi\overline{R}cc}$ was found to be sensitive to such a difference in implementation, producing an incorrect relative sign between diagrams with crossed external ghost lines.
We note that such diagrams with identical external fermions do not appear in the Abelian calculation.

\section{Results for the two-loop counterterms}\label{Sec:CountertermResults}
In this section we provide the explicit and full counterterm Lagrangian of both UV renormalization and restoration of symmetry breaking $\mathcal{L}^{2\mathrm{L}}_{\mathrm{ct}}=\mathcal{L}^{2\mathrm{L}}_{\mathrm{sct}}+\mathcal{L}^{2\mathrm{L}}_{\mathrm{fct}}$. The one-loop counterterms needed for subrenormalization are essentially the ones of Ref.~\cite{Belusca-Maito:2020ala} but have been re-derived by doing the complete one-loop renormalization in our setup adapted to the reducible $SU(2)$-model. 

The technical setup is similar to the ones used in the Abelian two-loop and three-loop calculations of Refs.~\cite{Belusca-Maito:2021lnk, Stockinger:2023ndm}. The  results for loop diagrams are obtained using the well-known tadpole expansion method \cite{Misiak:1994zw, Chetyrkin:1997fm, Lang:2020nnl}, by introducing an auxiliary mass into the denominators and setting external momenta to zero. The calculations were performed with the help of Mathematica, specifically the packages \texttt{FeynRules}, \texttt{FeynArts}, \texttt{FeynHelpers} and \texttt{FeynCalc} \cite{Alloul:2013bka, Hahn:2000kx, Shtabovenko:2016sxi, Shtabovenko:2020gxv, Shtabovenko:2021hjx, Shtabovenko:2016whf}. As discussed
in Sec.~\ref{SubSec:NASubtleties}, powerful consistency checks are possible by separately evaluating and comparing both the LHS and RHS of the quantum action principle (cf.~Eq.~(\ref{Eq:ExampleQAP3Pt})). These consistency checks were performed for the divergent parts of all breaking Green functions. Additionally we also performed the consistency checks for the finite parts for the Green functions $(\Delta\cdot\Gamma)_{Gc}$ and $(\Delta\cdot\Gamma)_{\psi\overline{\psi}c}$. The latter calculation employs \texttt{TARCER} \cite{Mertig:1990an} in order to evaluate the finite part of some of the two-loop integrals, providing further independent checks.
Further non-trivial consistency checks arise from the absence of logarithms in the two-loop counterterms required to cancel UV divergences and symmetry breakings
after subrenormalization and all subtleties mentioned in Sec.~\ref{SubSec:NASubtleties}
are taken into account.

\subsection{Two-loop divergent counterterms}

We provide the complete set of two-loop UV-divergent counterterms. They are determined by computing the divergent parts of all power-counting divergent Green functions with ghost number zero. The contributing Green functions are the same as at one-loop order in Refs.~(\cite{Martin:1999cc, Belusca-Maito:2020ala}).
Contrary to the one-loop results, we have refrained from arranging the pieces in explicitly gauge-invariant combinations due to the proliferation of non-symmetric terms. As one consistency check, the divergent counterterms determined via ordinary Green functions enter Eq.~(\ref{Eq:DeltaGammaAt2Loop}) via the last term and thus are primarily responsible for the cancellation of the UV divergences of the two-loop breaking Green functions with $\Delta$-insertions.

In the following we first list separately the result of the divergent two-loop counterterm Lagrangian for each field combination, suppressing a global factor of $-\frac{1}{256\pi^4}$. Together they will sum up to $\mathcal{L}^{2\mathrm{L}}_{\mathrm{sct}}$. We introduce the compact notation,
\begin{align}
    G\,\square\,G&\equiv G^{a\mu}\partial^{\nu}\partial^{\nu}G^{a\mu}
    &(\partial\widetilde{G)G)G}&\equiv\varepsilon^{abc}(\partial^{\mu}G^{a\nu})G^{b\nu}G^{c\mu},\nonumber\\
    G\partial\partial G&\equiv G^{a\mu}\partial^{\mu}\partial^{\nu}G^{a\nu}
    &G^2&\equiv G^{a\mu}G^{a\mu},\nonumber\\
C_2(\mathcal{R})\overline{\psi}\slashed{\partial}\psi&\equiv\overline{\psi}_{i}C_2(\mathcal{R})_{ij}\slashed{\partial}\psi_{j}
    &G^4&\equiv (G^2)^2,\\
    \overline{\psi}T_{\mathcal{R}}\slashed{G}\psi&\equiv\overline{\psi}_{i}T^a_{\mathcal{R}ij}\slashed{G}^a\psi_{j}
    &(GG)(GG)&\equiv G^{a\mu}G^{b\mu}G^{a\nu}G^{b\nu}.\nonumber
\end{align}

\paragraph{Gauge Boson Self-Energy}

\begin{equation}
    \begin{aligned}
        &\phantom{+}g^4\biggl(\frac{25}{6\epsilon^2}-\frac{23}{4\epsilon}\biggr)
        \Bigl(G\,\square\,G-G\partial\partial G\Bigr)+g^4\sum_{\mathcal{R}}\bigl(S_2(\mathcal{R})\bigr)\biggl(-\frac{5}{6\epsilon^2}\Bigl(\overline{G\,\square\,G}-\overline{G\partial\partial G}\Bigr)\\
        &+\frac{23}{18\epsilon}\Bigl(\overline{G\,\square\,G}-\overline{G\partial\partial G}\Bigr)-\frac{17}{36\epsilon}\overline{G\partial\partial G}
        +\biggl(\frac{5}{4\epsilon^2}-\frac{371}{144\epsilon}\biggr)\overline{G\partial}\widehat{\partial G}\\
        &+
        \biggl(-\frac{7}{8\epsilon^2}+\frac{43}{288\epsilon}\biggr)\overline{G}\,\widehat{\square}\,\overline{G}
        +\biggl(-\frac{13}{18\epsilon^2}+\frac{103}{54\epsilon}\biggr)\widehat{G}\,\overline{\square}\,\widehat{G}\\
        &+
        \biggl(\frac{11}{36\epsilon^2}-\frac{317}{432\epsilon}\biggr)\widehat{G\,\square\,G}
        +\biggl(-\frac{1}{3\epsilon^2}+\frac{13}{18\epsilon}\biggr)\widehat{G\partial\partial G}\biggr)\\
        &+
        g^4\sum_{\mathcal{R}}\bigl(S_2(\mathcal{R})C_2(\mathcal{R})\bigr)
        \biggl(
        \frac{1}{3\epsilon}\Bigl(\overline{G\,\square\,G}-\overline{G\partial\partial G}\Bigr)
        +\biggl(\frac{1}{12\epsilon^2}-\frac{17}{144\epsilon}\biggr)\overline{G}\,\widehat{\square}\,\overline{G}
        \biggr)
    \end{aligned}
\end{equation}

\paragraph{Fermion Self-Energy}

\begin{equation}
    \begin{aligned}
       &\phantom{+}
        ig^4\biggl(\biggl(-\frac{1}{2\epsilon^2}-\frac{7}{12\epsilon}\biggr)C_2(\mathcal{R})
        -\frac{2}{\epsilon^2}+\frac{107}{18\epsilon}-\frac{1}{9\epsilon}\sum_{\mathcal{R}}\bigl(S_2(\mathcal{R})\bigr)\biggr)\overline{\psi}C_2(\mathcal{R})\overline{\slashed{\partial}}\projR\psi
    \end{aligned}
\end{equation}

\paragraph{Ghost Self-Energy}

\begin{equation}
    \begin{aligned}
        &\phantom{+}
        g^4\biggl(-\frac{4}{\epsilon^2}+\frac{49}{12\epsilon}\biggr)\overline{c}\,\square\,c+g^4\sum_{\mathcal{R}}\bigl(S_2(\mathcal{R})\bigr)\biggl(\frac{1}{2\epsilon^2}-\frac{23}{18\epsilon}\biggr)\overline{c}\,\overline{\square}\,c
    \end{aligned}
\end{equation}

\paragraph{$\rho$-Ghost}

\begin{equation}
    \begin{aligned}
        &\phantom{+}
        g^4\biggl(\frac{4}{\epsilon^2}-\frac{49}{12\epsilon}\biggr)\rho\partial c+g^4\sum_{\mathcal{R}}\bigl(S_2(\mathcal{R})\bigr)\biggl(-\frac{1}{2\epsilon^2}+\frac{23}{18\epsilon}\biggr)\overline{\rho\partial} c
    \end{aligned}
\end{equation}

\paragraph{$\overline{R}$-Ghost-Fermion}

\begin{equation}
    \begin{aligned}
        &\phantom{+}
        ig^5\biggl(-\frac{5}{2\epsilon^2}+\frac{1}{9\epsilon}\sum_{\mathcal{R}}\bigl(S_2(\mathcal{R})\bigr)
        -\frac{1}{3\epsilon}C_2(\mathcal{R})\biggr)
        \overline{R}cT_{\mathcal{R}}\projR\psi
    \end{aligned}
\end{equation}

\paragraph{$R$-Ghost-Fermion}

\begin{equation}
    \begin{aligned}
       &\phantom{+}
        ig^5\biggl(-\frac{5}{2\epsilon^2}+\frac{1}{9\epsilon}\sum_{\mathcal{R}}\bigl(S_2(\mathcal{R})\bigr)
        -\frac{1}{3\epsilon}C_2(\mathcal{R})
        \biggr)\overline{\psi}cT_{\mathcal{R}}\projL R
    \end{aligned}
\end{equation}

\paragraph{Fermion-Gauge Boson Vertex}

\begin{equation}
    \begin{aligned}
        \phantom{+}
        &g^5\biggl(
        \biggl(-\frac{1}{2\epsilon^2}-\frac{17}{12\epsilon}\biggr)(C_2(\mathcal{R}))^2
        +
        \biggl(-\frac{4}{\epsilon^2}+\frac{20}{9\epsilon}\biggr)C_2(\mathcal{R})
        +
        \frac{1}{9\epsilon}C_2(\mathcal{R})\sum_{\mathcal{R}}\bigl(S_2(\mathcal{R})\bigr)\\
        &+
        \biggl(\frac{1}{2\epsilon^2}-\frac{16}{9\epsilon}\biggr)\sum_{\mathcal{R}}\bigl(S_2(\mathcal{R})\bigr)
        +
        \biggl(-\frac{17}{2\epsilon^2}+\frac{67}{12\epsilon}\biggr)
        \biggr)
        \overline{\psi}T_{\mathcal{R}}\overline{\slashed{G}}\projR\psi
    \end{aligned}
\end{equation}

\paragraph{Triple Gauge Boson Vertex}

\begin{equation}
    \begin{aligned}
    &g^5\biggl(\frac{13}{2\epsilon^2}-\frac{71}{12\epsilon}\biggr)\bigl(\bigl(\partial\widetilde{G\bigr)G\bigr)G}+g^5\frac{1}{3\epsilon}\sum_{\mathcal{R}}\bigl(S_2(\mathcal{R})C_2(\mathcal{R})\bigr)\bigl(\overline{\partial}\widetilde{\overline{G}\bigr)\overline{G}\bigr)\overline{G}}\\
      &+g^5\sum_{\mathcal{R}}\bigl(S_2(\mathcal{R})\bigr)\biggl(
      \biggl(-\frac{5}{2\epsilon^2}+\frac{67}{144\epsilon}\biggr)\bigl(\bigl(\overline{\partial}\widetilde{\overline{G}\bigr)\overline{G})\overline{G}}+
      \biggl(-\frac{91}{96\epsilon^2}+\frac{2477}{1152\epsilon}\biggr)\bigl(\bigl(\overline{\partial}\widetilde{\widehat{G}\bigr)\widehat{G})\overline{G}}\\
      &+
\biggl(-\frac{65}{96\epsilon^2}+\frac{619}{1152\epsilon}\biggr)\bigl(\bigl(\widehat{\partial}\widetilde{\overline{G}\bigr)\overline{G})\widehat{G}}
        +
        \biggl(\frac{7}{8\epsilon^2}-\frac{347}{288\epsilon}\biggr)\bigl(\bigl(\widehat{\partial}\widetilde{\widehat{G}\bigr)\widehat{G})\widehat{G}})
      \biggr)
    \end{aligned}
\end{equation}

\paragraph{$\rho$-Ghost-Gauge Boson}

\begin{equation}
    \begin{aligned}
        g^5\biggl(-\frac{5}{2\epsilon^2}+\frac{3}{2\epsilon}\biggr)\widetilde{\rho G c}+g^5\frac{1}{9\epsilon}\sum_{\mathcal{R}}\bigl(S_2(\mathcal{R})\bigr)\widetilde{\widehat{\rho} \widehat{G} c}-g^5\frac{1}{12\epsilon}\sum_{\mathcal{R}}\bigl(S_2(\mathcal{R})\bigr)\widetilde{\overline{\rho} \overline{G} c}
    \end{aligned}
\end{equation}

\paragraph{$\zeta$-Ghost-Ghost}

\begin{equation}
    \begin{aligned}
        g^5\biggl(\frac{5}{4\epsilon^2}-\frac{3}{4\epsilon}-\frac{1}{18\epsilon}\sum_{\mathcal{R}}\bigl(S_2(\mathcal{R})\bigr)\biggr)\widetilde{\zeta cc}
    \end{aligned}
\end{equation}

\paragraph{Ghost-Gauge Boson Vertex}

\begin{equation}
    \begin{aligned}
        g^5\biggl(-\frac{5}{2\epsilon^2}+\frac{3}{2\epsilon}\biggr)\widetilde{\bigl(\partial\overline{c}\bigr)Gc}+
        g^5\frac{1}{9\epsilon}\sum_{\mathcal{R}}\bigl(S_2(\mathcal{R})\bigr)\widetilde{\bigl(\widehat{\partial}\overline{c}\bigr)\widehat{G}c}-
        g^5\frac{1}{12\epsilon}\sum_{\mathcal{R}}\bigl(S_2(\mathcal{R})\bigr)\widetilde{\bigl(\overline{\partial}\overline{c}\bigr)\overline{G}c}
    \end{aligned}
\end{equation}

\paragraph{Quartic Gauge Boson Vertex}

\begin{equation}
    \begin{aligned}
        &g^6\biggl(-\frac{1}{6\epsilon^2}+\frac{1}{12\epsilon}\biggr)\Bigl(G^4-\bigl(GG\bigr)\bigl(GG\bigr)\Bigr)
        +g^6\sum_{\mathcal{R}}\bigl(S_2(\mathcal{R})\bigr)\biggl(\frac{5}{6\epsilon^2}\Bigl(\overline{G}^4-\bigl(\overline{GG}\bigr)\bigl(\overline{GG}\bigr)\Bigr)\\
        &+\frac{1249}{2880\epsilon}\overline{G}^4-\frac{1897}{2880\epsilon}\bigl(\overline{GG}\bigr)\bigl(\overline{GG}\bigr)
        +\biggl(\frac{19}{72\epsilon^2}-\frac{2377}{8640\epsilon}\biggr)\overline{G}^2\widehat{G}^2
        +\biggl(-\frac{73}{288\epsilon^2}+\frac{919}{3456\epsilon}\biggr)\widehat{G}^4\\
        &+\biggl(-\frac{43}{288\epsilon^2}+\frac{827}{17280\epsilon}\biggr)\bigl(\overline{GG}\bigr)\bigl(\widehat{GG}\bigr)
        +\biggl(\frac{95}{288\epsilon^2}-\frac{641}{3456\epsilon}\biggr)\bigl(\widehat{GG}\bigr)\bigl(\widehat{GG}\bigr)\biggr)\\
        &+g^6\frac{1}{\epsilon}\sum_{\mathcal{R}}\bigl(S_2(\mathcal{R})C_2(\mathcal{R})\bigr)\biggl(\frac{11}{60}\overline{G}^4+\frac{1}{15}\overline{G}^2\widehat{G}^2+\frac{7}{60}\bigl(\overline{GG}\bigr)\bigl(\overline{GG}\bigr)-\frac{1}{30}\bigl(\overline{GG}\bigr)\bigl(\widehat{GG}\bigr)\biggr)
    \end{aligned}
\end{equation}

Taken together the preceding results combine to give the full divergent two-loop counterterm Lagrangian,
\begin{equation}
\begin{aligned}
    \mathcal{L}^{2\mathrm{L}}_{\mathrm{sct}}=-\frac{1}{256\pi^4}&\bigl((4.2)+(4.3)+(4.4)+(4.5)+(4.6)\\&+(4.7)+(4.8)+(4.9)+(4.10)+(4.11)+(4.12)+(4.13)\bigr).
\end{aligned}
\end{equation}
Supplementing the one-loop renormalized action with these two-loop divergent counterterms ensures that all Green functions up to the two-loop level become finite. This includes the finiteness of the symmetry breaking in Eq.~(\ref{Eq:DeltaGammaAt2Loop}).

\subsection{Two-loop finite counterterms}

The finite counterterms $S_{\mathrm{fct}}^{2\mathrm{L}}$ are determined as described in Sec.~\ref{Sec:DeterminationOfFCT}. In practice we set up an ansatz of all possible $4$-dimensional (i.e.~non-evanescent) monomials of mass dimension $\leq4$ and ghost number $=0$ with coefficients to be determined. We compute the action of $b_D$ (cf.~Eq.~(\ref{Eq:bDDef})) on the ansatz to obtain $\Delta^{2\mathrm{L}}_{\mathrm{fct}}$, defined in Eq.~(\ref{Eq:DeltaAt2Loop}) and corresponding to the last term in Eq.~(\ref{Eq:DeltaGammaAt2Loop}). As  described in Sec.~\ref{Sec:DeterminationOfFCT}, the other terms in that equation  correspond to genuine two-loop diagrams with tree-level vertex $\widehat{\Delta}$, one-loop diagrams with $\Delta^1_{\mathrm{ct}}$-insertions and products of counterterms without $\Delta$. They all have to be computed, but the computation is unambiguous given  results from tree-level, from one-loop renormalization and  two-loop UV renormalization.

The list of $\widehat{\Delta}$-Green functions that need to be computed 
is given by the list of all power-counting divergent dimension $\leq4$, ghost-number $=1$ Green functions with a single insertion of $\widehat{\Delta}$. They comprise those of Refs.~\cite{Belusca-Maito:2020ala, Martin:1999cc} as well as new Green functions discussed in Sec.~\ref{Sec:DeterminationOfFCT}, which happened to vanish at one-loop order. Green functions with external antighost always have reduced UV power counting at any loop order and may thus be neglected. The external antighost line can only connect via the $\overline{c}Gc$-vertex, which carries the external antighost momentum.
Thus Green functions such as $(\Delta\cdot\Gamma)_{GG\overline{c}c}$ and $(\Delta\cdot\Gamma)_{c c\overline{c}\overline{c}c}$ are immediately seen to have vanishing finite part in Eq.~(\ref{Eq:STIRenCond}). The only potentially non-trivial five-point function is $(\Delta\cdot\Gamma)_{GGGGc}$.

With these results we get a generally overdetermined system of equations for the coefficients in the counterterm ansatz. Because of non-linear BRST transformations, monomials such as $GG$ in $S_{\mathrm{fct}}^{2\mathrm{L}}$ contribute to different breaking Green functions, $(\Delta\cdot\Gamma)_{Gc}$ and $(\Delta\cdot\Gamma)_{GGc}$. In addition, counterterms with external sources typically transform into many different breaking operators, such as the monomial $\rho c G$ in $S_{\mathrm{fct}}^{2\mathrm{L}}$ whose $b_D$-transformation may contribute to
$(\Delta\cdot\Gamma)_{GGc}$, $(\Delta\cdot\Gamma)_{GGGc}$, $(\Delta\cdot\Gamma)_{GGGGc}$, $(\Delta\cdot\Gamma)_{Gc\overline{c}c}$, $(\Delta\cdot\Gamma)_{G c\rho c}$ and $(\Delta\cdot\Gamma)_{G\psi\overline{\psi}c}$.
Hence there exist non-trivial consistency relations among breaking Green functions that must be respected for a solution for the finite counterterm action to exist at all. Still there is some freedom in choosing the solution as finite, symmetric counterterms can always be added to $S_{\mathrm{fct}}$ without spoiling Eq.~(\ref{Eq:STIRenCond}). As an example, by adding the gauge-invariant counterterm $\overline{F}^2$, we could alter the coefficients of the counterterm monomials $S^{2\mathrm{L}}_{\mathrm{fct},\,GG}$, $S^{2\mathrm{L}}_{\mathrm{fct},\,GGG}$ and $S^{2\mathrm{L}}_{\mathrm{fct},\,GGGG}$.

We obtain a solution for the resulting system of equations, which verifies the validity of the necessary consistency conditions.
Since the solution is not unique, we have to fix a solution for the coefficients by choosing to recover a similar counterterm structure as at one-loop (cf.~Ref.~\cite{Belusca-Maito:2020ala}). Indeed we manage to find the same set of counterterm monomials as at one-loop, including the external sources $\overline{R}$ and $R$, as well as a novel counterterm to the external source $\zeta$. A new kind of counterterm to either $\zeta$ or $\rho$ is found to be unavoidable for any solution.

In the following we list the complete solution $\mathcal{L}^{2\mathrm{L}}_{\mathrm{fct}}$ for the finite symmetry-restoring counterterms, first term by term, then in combination. The notation is the same as for the divergent counterterms, and a global factor of $-\frac{1}{256\pi^4}$ is suppressed.

\paragraph{Gauge Boson Self-Energy}

\begin{equation}
    \begin{aligned}
        g^4\biggl(-\frac{11}{48}\sum_{\mathcal{R}}\bigl(C_2(\mathcal{R})S_2(\mathcal{R})\bigr)+\frac{7}{54}\sum_{\mathcal{R}}\bigl(S_2(\mathcal{R})\bigr)\biggr)\overline{G\,\square\,G}
    \end{aligned}
\end{equation}

\paragraph{Fermion Self-Energy}

\begin{equation}
    \begin{aligned}
        ig^4\biggl(\frac{127}{36}\bigl(C_2(\mathcal{R})\bigr)^2&+\frac{41}{108}C_2(\mathcal{R})\sum_{\mathcal{R}}\bigl(S_2(\mathcal{R})\bigr)\\
        &-\frac{25}{72}\sum_{\mathcal{R}}\bigl(S_2(\mathcal{R})\bigr)
        -\frac{1181}{108}C_2(\mathcal{R})
        +\frac{31}{6}\biggr)\overline{\psi}\overline{\slashed{\partial}}\projR\psi
    \end{aligned}
\end{equation}

\paragraph{$\overline{R}$-Ghost-Fermion}

\begin{equation}
    \begin{aligned}
        ig^5\biggl(\frac{1}{18}C_2(\mathcal{R})-\frac{35}{216}\sum_{\mathcal{R}}\bigl(S_2(\mathcal{R})\bigr)+\frac{1}{6}\biggr)\overline{R}cT_{\mathcal{R}}\projR\psi
    \end{aligned}
\end{equation}

\paragraph{$R$-Fermion-Ghost}

\begin{equation}
    \begin{aligned}
        ig^5\biggl(\frac{1}{18}C_2(\mathcal{R})-\frac{35}{216}\sum_{\mathcal{R}}\bigl(S_2(\mathcal{R})\bigr)+\frac{1}{6}\biggr)\overline{\psi}cT_{\mathcal{R}}\projL cR
    \end{aligned}
\end{equation}

\paragraph{Triple Gauge Boson Vertex}

\begin{equation}
    \begin{aligned}
        -g^5\biggl(\frac{23}{72}\sum_{\mathcal{R}}\bigl(C_2(\mathcal{R})S_2(\mathcal{R})\bigr)+\frac{365}{1728}\sum_{\mathcal{R}}\bigl(S_2(\mathcal{R})\bigr)\biggr)\bigl(\bigl(\overline{\partial}\widetilde{\overline{G}\bigr)\overline{G}\bigr)\overline{G}}
    \end{aligned}
\end{equation}

\paragraph{$\zeta$-Ghost-Ghost}

\begin{equation}
    \begin{aligned}
        g^5\frac{35}{432}\sum_{\mathcal{R}}\bigl(S_2(\mathcal{R})\bigr)\widetilde{\zeta cc}
    \end{aligned}
\end{equation}

\paragraph{Quartic Gauge Boson Vertex}

\begin{equation}
    \begin{aligned}
        &g^6\biggl(\frac{3}{80}\sum_{\mathcal{R}}\bigl(\bigl(C_2(\mathcal{R})\bigr)^2S_2(\mathcal{R})\bigr)+\frac{31}{720}\sum_{\mathcal{R}}\bigl(C_2(\mathcal{R})S_2(\mathcal{R})\bigr)+\frac{137}{768}\sum_{\mathcal{R}}\bigl(S_2(\mathcal{R})\bigr)\biggr)\overline{G}^4\\
        &+g^6\biggl(\frac{3}{40}\sum_{\mathcal{R}}\bigl(\bigl(C_2(\mathcal{R})\bigr)^2S_2(\mathcal{R})\bigr)-\frac{7}{180}\sum_{\mathcal{R}}\bigl(C_2(\mathcal{R})S_2(\mathcal{R})\bigr)
        -\frac{403}{2304}\sum_{\mathcal{R}}\bigl(S_2(\mathcal{R})\bigr)\biggr)\bigl(\overline{GG}\bigr)\bigl(\overline{GG}\bigr)
    \end{aligned}
\end{equation}

Taken together the preceding results combine to give the full finite, $4$-dimensional two-loop counterterm Lagrangian,
\begin{equation}
    \mathcal{L}^{2\mathrm{L}}_{\mathrm{fct}}=-\frac{1}{256\pi^4}\bigl((4.15)+(4.16)+(4.17)+(4.18)+(4.19)+(4.20)+(4.21)\bigr).
\end{equation}
After adding these counterterms to the two-loop action, the symmetry relations of Eq.~(\ref{Eq:STIRenCond}) are fulfilled in the limit $\mathrm{LIM}_{D\rightarrow 4}$.

\section{Conclusion}\label{Sec:Conclusions}
In this paper we have presented the first complete two-loop renormalization of a non-Abelian gauge theory in the BMHV scheme. The result comprises two kinds of two-loop counterterms. The divergent counterterms render the theory finite; they involve symmetric and non-symmetric parts, and they involve non-evanescent and evanescent terms. The finite counterterms render the renormalized theory symmetric, i.e.\ they restore the validity of the Slavnov-Taylor identity. The finite counterterms are purely non-evanescent. They could be further modified by adding finite symmetric counterterms (arising e.g.\ from field and parameter renormalization) to satisfy desired renormalization conditions.

Compared to the one-loop level the finite counterterm action is structurally very similar and hence just as compact. In particular, it contains the same counterterm monomials as the one-loop result except for an additional finite breaking contributions involving the external source $\zeta$.

We have also discussed challenges encountered in non-Abelian theories at the multi-loop level. The proliferation of terms involving external source Green functions and the effects of their subrenormalization reveal the whole range of technical pitfalls contained in Eq.~(\ref{Eq:DeltaGammaAt2Loop}) that are nevertheless indispensable for a consistent renormalization.
We successfully implemented a \texttt{FeynArts}-model file including the exotic $\widehat{\Delta}$-vertex as well as external source operators, which allows to generate the amplitudes of all diagrams in an automated fashion. In particular it managed to reproduce the known one-loop results. For the two-loop calculation we had to handle a new kind of vertex structure, composed both of a $\Delta$-insertion and an external source.

As a general prospect our result demonstrates that the successful application of the all-order consistent BMHV scheme is practically feasible for non-Abelian chiral gauge theories even on the multi-loop level. In particular the method advanced in our previous papers, which relies on the quantum action principle and evanescent operator insertions, is reaffirmed as a powerful approach to systematically determining the finite counterterms despite cumbersome computational artifacts of the scheme.

The insights and setup can be utilized in upcoming (multi-)loop calculations in more realistic models. For example the Lagrangian could be supplemented with scalars (cf.~\cite{Belusca-Maito:2020ala,OlgosoRuiz:2024dzq,Ebert:2024xpy}) as well as interacting left-handed fermions (cf.~\cite{Cornella:2022hkc,OlgosoRuiz:2024dzq, Ebert:2024xpy}) to renormalize the SM in the BMHV scheme at the two-loop level.

\section*{Acknowledgments} 
P.K. and D.S.\ acknowledge financial support by the German Science 
Foundation DFG, grant STO 876/8-1 and STO 876/8-2.
We would like to thank Matthias Weißwange and Amon Ilakovac for insightful ideas 
and valuable discussions, as well as Baibhab Ray for initial collaboration.

\printbibliography

\end{document}